\documentclass[10pt]{article}
\usepackage{graphicx}
\def\Title#1{\begin{center} {\Large #1 } \end{center}}
\def\Author#1{\begin{center}{ \sc #1} \end{center}}
\def\Address#1{\begin{center}{ \it #1} \end{center}}

\newcommand\pubblock{\rightline{\begin{tabular}{l} Proceedings of the Fifth Annual LHCP\\ \pubnumber\\
         \pubdate  \end{tabular}}}

\newenvironment{Abstract}{\begin{quotation} \begin{center} 
             \large ABSTRACT \end{center}\bigskip 
      \begin{large}}{\end{large} \end{quotation}}

\newenvironment{Presented}{\begin{quotation} \begin{center} 
             PRESENTED AT\end{center}\bigskip 
      \begin{center}\begin{large}}{\end{large}\end{center} \end{quotation}}

\def\beq{\begin{equation}}
\def\eeq#1{\label{#1}\end{equation}}
\def\eeqn{\end{equation}}

\def\beqa{\begin{eqnarray}}
\def\eeqa#1{\label{#1}\end{eqnarray}}
\def\eeqan{\end{eqnarray}}

\let\bar=\overbar

\def\Dslash{\not{\hbox{\kern-4pt $D$}}}
\def\dslash{\not{\hbox{\kern-2pt $\del$}}}

\def\msb{{\bar{\ssstyle M \kern -1pt S}}}

\usepackage{color}

\newcommand\pubnumber{ CMS CR-2017/322 }

\newcommand\pubdate{\today}

\def\affiliation{
(On behalf of the ATLAS \& CMS Collaborations) \\
School of Physical Sciences \\
National Institute of Science Education and Research\\
Bhimpur-Padanpur (Jatni), Dist Khurda, Odisha 752050, India}

\begin{document}

\large
\begin{titlepage}
\pubblock

\vfill
\Title{ Top Quark Decay Properties }
\vfill

\Author{ Prolay Kumar Mal }
\Address{\affiliation}
\vfill
\begin{Abstract}
Due to the large production cross-section, many of the top quark properties can be
measured very precisely at the LHC. A very few recent results, probed only through the
top quark decay vertices are presented here. These results are based on proton-proton
collision datasets recorded by the ATLAS and CMS experiments at $\sqrt{s}$=7, 8 and
13 TeV. All the measurements and observed limits are consistent with the
Standard Model (SM) predictions, while strong bounds on anomalous Wtb couplings are
established.
\end{Abstract}
\vfill

\begin{Presented}
The Fifth Annual Conference\\
 on Large Hadron Collider Physics \\
Shanghai Jiao Tong University, Shanghai, China\\ 
May 15-20, 2017
\end{Presented}
\vfill
\end{titlepage}
\def\thefootnote{\fnsymbol{footnote}}
\setcounter{footnote}{2}
%

\normalsize 


\section{Introduction}
Measurement of top quark properties is one of unique ways 
to scrutinize the Standard Model (SM) predictions thoroughly,
along with the possibility to probe new physics signature. At
the LHC, the top quarks are produced through strong and
electroweak processes at an unprecedented rate \cite{ref_topprod} and such
a large statistics of top quark events provides the opportunity
to explore both production and decay vertices. While the
measurements directly related to the production mechanism are
reported in a separate article (in this set of conference proceedings),
this proceeding focuses only on the decay properties covering 
only a few recent results from the ATLAS\cite{ref_atlasdet} and CMS\cite{ref_cmsdet} experiments:
W-boson helicity measurement, top quark decay width, top quark
branching fraction (considering only the Flavor Changing Charge Current
decays\footnote{Flavor Changing Neutral Current decays or FCNC and
other rare decays are discussed in another article in this set of proceedings})
and anomalous Wtb coupling. These results are based on the proton-proton
collision datasets at $\sqrt{s}$=7, 8 and 13 TeV recorded during 2011,
2012 and 2016 LHC operations respectively.

\section{W-boson helicity}\label{sec:whelicity}
Within the SM the top quarks decays into a W-boson and a b-quark with almost 100\%
branching fraction and the polarization of the W-boson can be measured using the
$\rm t\bar{t}$ events. The W-boson helicity fractions (left-handed, right-handed
and longitudinal) are defined as $\rm F_{L,R,0} = \Gamma_{L,R,0}/\Gamma_{Total}$, where
$\rm \Gamma_{L,R,0}$ are the partial decay widths in left-handed, right-handed, and
longitudinal helicity states respectively, with $\rm \Gamma_{Total}$ being the
total decay width. The SM next-tonext-to-leading order (NNLO) calculations
\cite{ref_whelicitysm} including the electroweak effects predict the values of
$\rm F_L = 0.311\pm 0.005$, $\rm F_R = 0.0017\pm 0.0001$ and $\rm F_0 = 0.687\pm 0.005$,
for a top quark mass of $\rm 172.8\pm 1.3$ GeV. Experimentally, the helicity angle
$\rm\theta^*$ can be defined as the angle between the direction of either the
down-type quark or the charged lepton arising from the W-boson decay and the reversed
direction of the top quark, both in the rest frame of the W-boson. The differential
cross-sections as a function of $\rm\cos\theta^*$ can then be written as
\begin{eqnarray*}
\rm \frac{1}{\sigma}\frac{d\sigma}{d\cos\theta^*} = \frac{3}{8} (1-\cos\theta^*)^2 F_L 
+\frac{3}{8} (1+\cos\theta^*)^2 F_R
+\frac{3}{4} (\sin\theta^*)^2 F_0.
\end{eqnarray*}

\begin{figure}[htb]
\begin{center}
\includegraphics[width=2.5in]{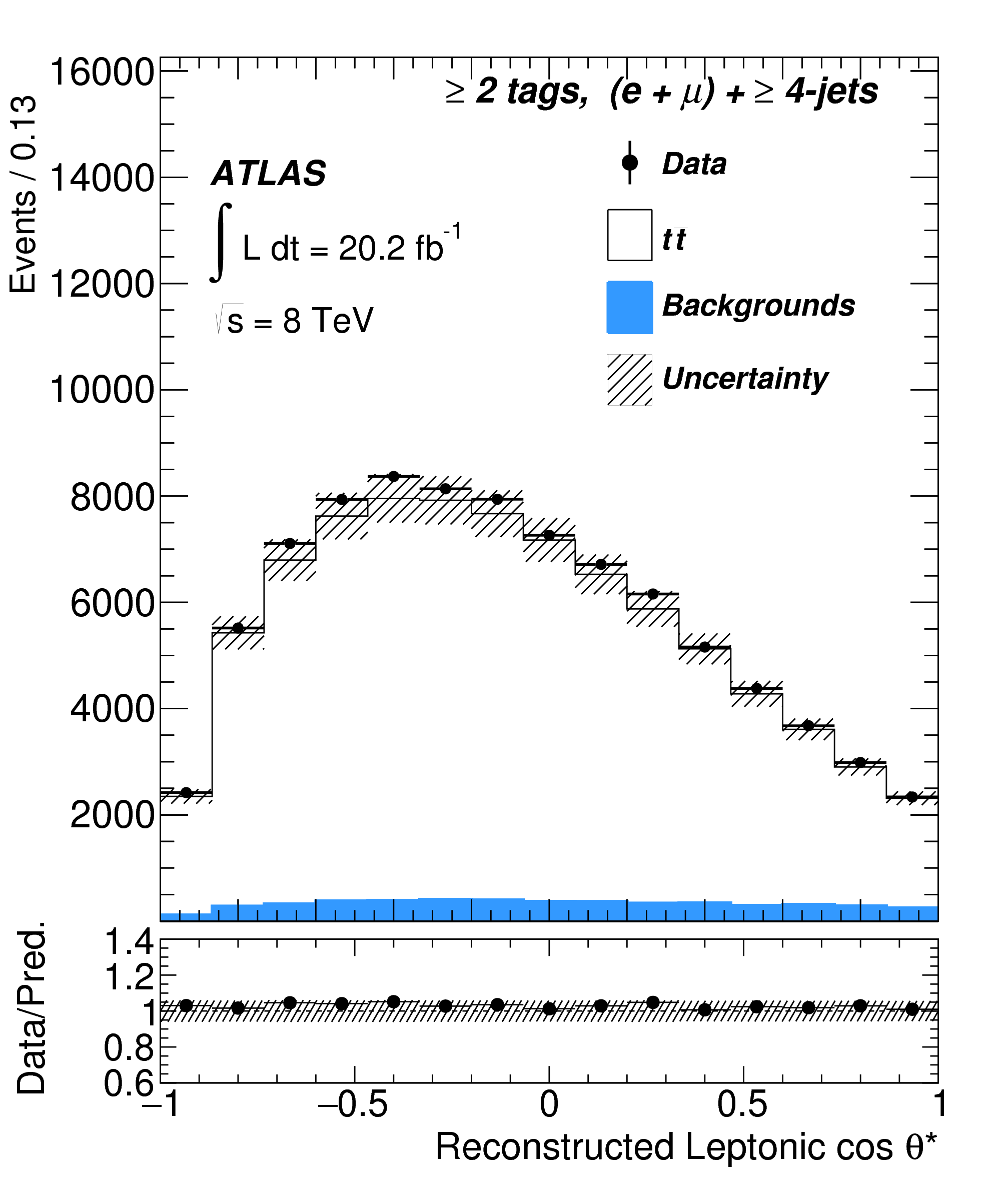}
\includegraphics[width=2.5in]{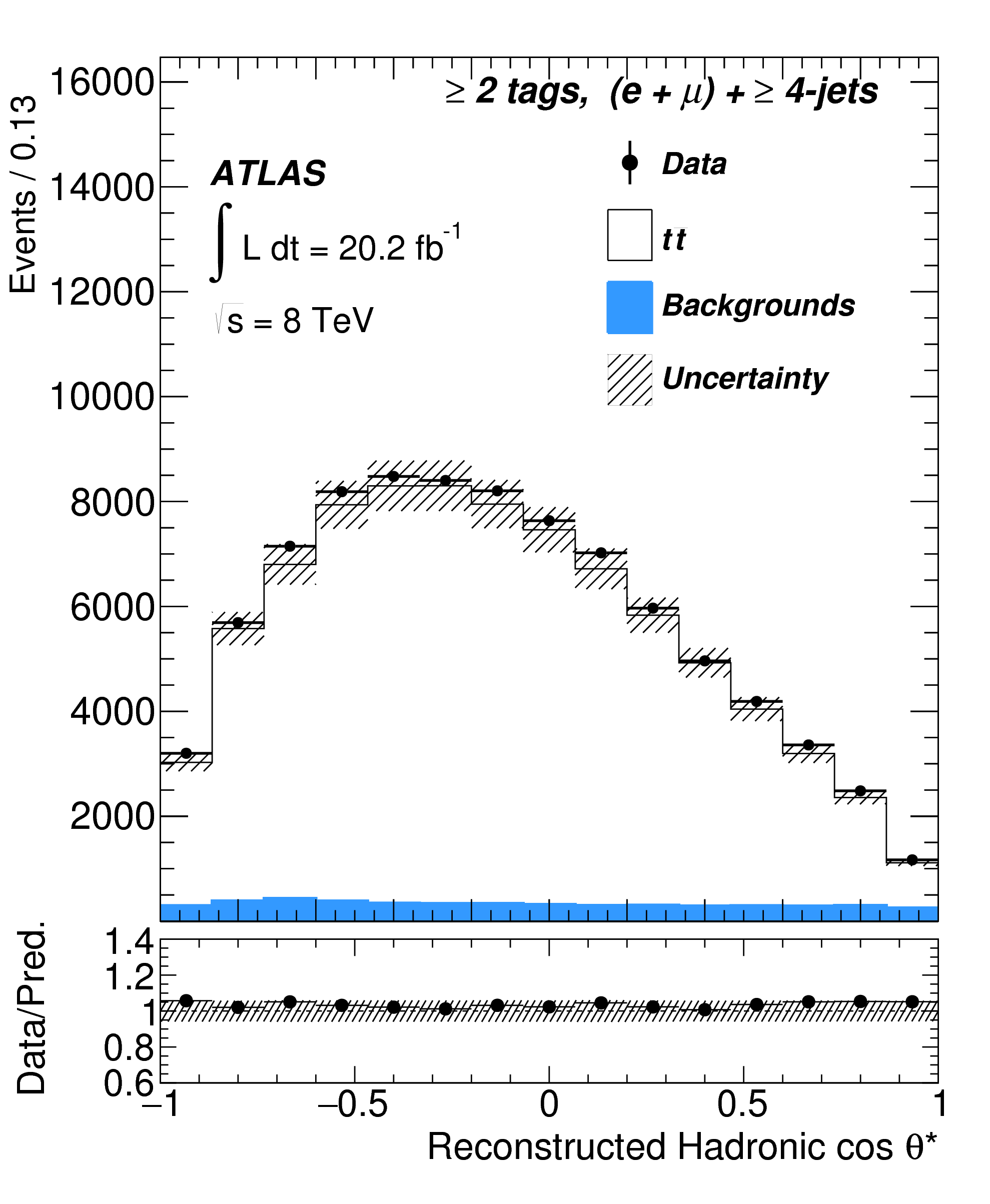}\\
\includegraphics[width=2.5in]{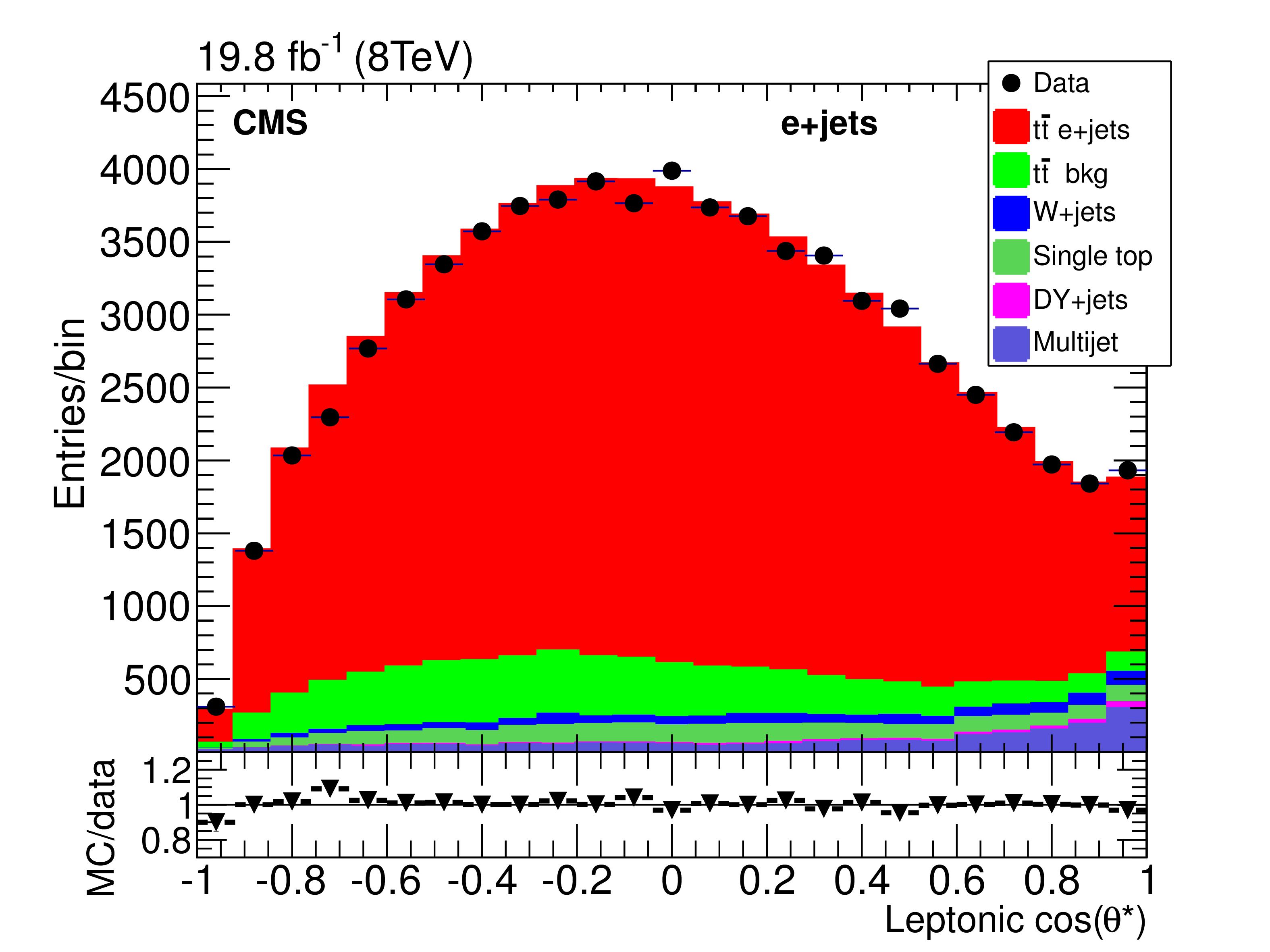}
\includegraphics[width=2.5in]{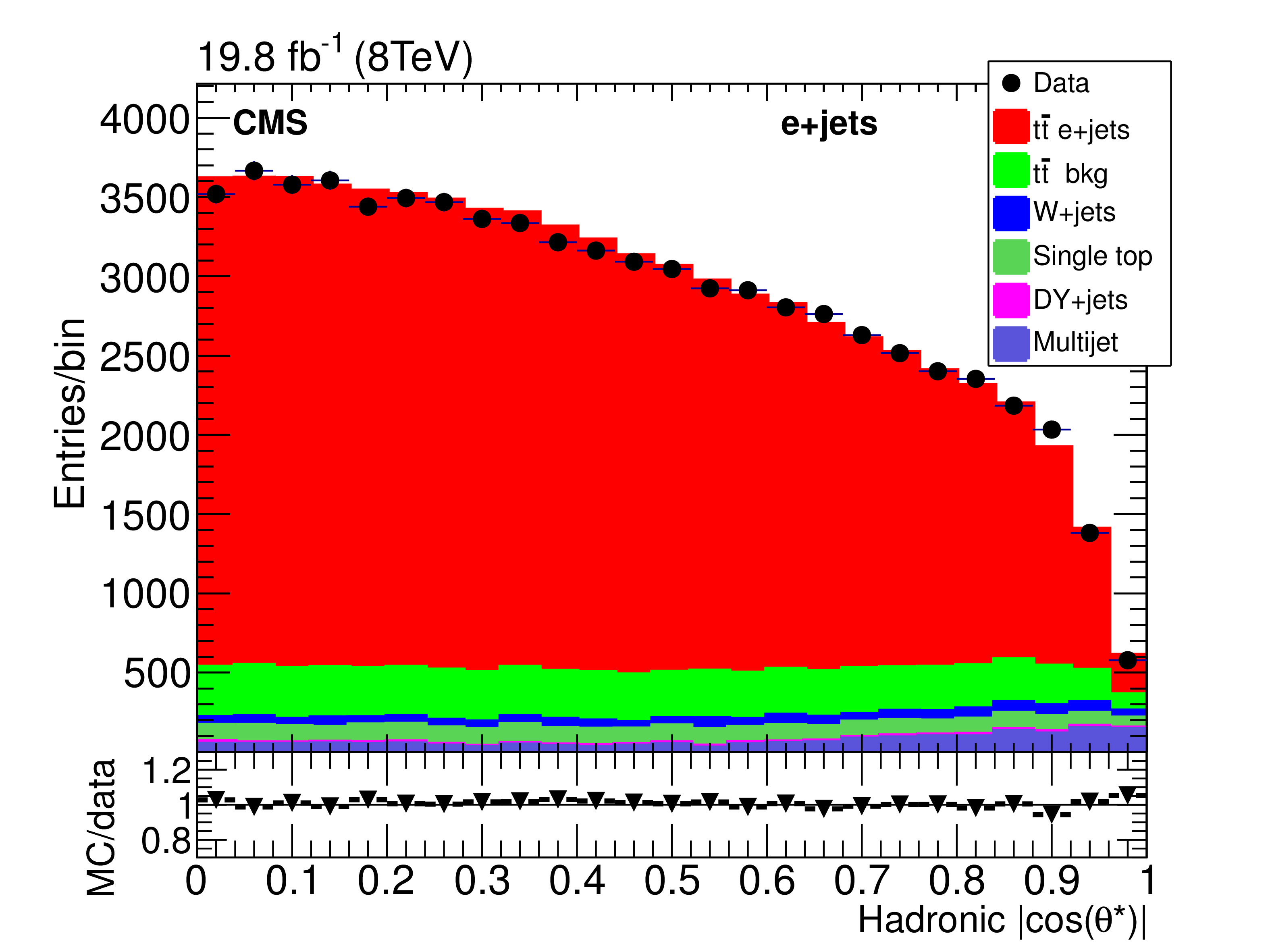}
\end{center}
\caption{Top: the leptonic (left) and hadronic (right) $\rm \cos\theta^*$
distributions\cite{ref_whelicityatlas} in
$\rm t\bar{t}\rightarrow$ lepton+jets events from ATLAS with $\rm\geq 2$
b-tagged jets; ATLAS datasets are sub-divided into four channels depending
on the lepton flavor and the b-jet multiplicity.
Bottom: inclusive leptonic (left) and hadronic (right)
$\rm \cos\theta^*$ distributions\cite{ref_whelicitycms} in $\rm t\bar{t}\rightarrow$ e+jets and
$\rm t\bar{t}\rightarrow \mu$+jets events from CMS.}
\label{fig:whelicity}
\end{figure}

ATLAS (CMS) have performed the measurements using $\rm t\bar{t}\rightarrow$ lepton+jets
events based on 20.2 $\rm fb^{-1}$ (19.8 $\rm fb^{-1}$) dataset from 2012 LHC operations.
The observables for W-boson helicity fractions i.e., `$\rm \cos\theta^*$'
are reconstructed using the leptonic ($\rm t\rightarrow bW\rightarrow bl\nu$) and hadronic
($\rm t\rightarrow bW\rightarrow bq\bar{q}'$) decay branches of the top quark, and
the distributions are shown in Fig.~\ref{fig:whelicity}. The CMS measurements 
\cite{ref_whelicitycms}
result in $\rm F_L = 0.323\pm 0.008 (stat)\pm 0.014 (syst)$,
$\rm F_R = 0.004\pm 0.005 (stat)\pm 0.014 (syst)$ and 
$\rm F_0 = 0.681\pm 0.012 (stat)\pm 0.023 (syst)$, while the ATLAS
measurements\cite{ref_whelicityatlas} quote
$\rm F_L = 0.299\pm 0.015$, $\rm F_R = 0.008\pm 0.014$ and
$\rm F_0 = 0.709\pm 0.019$. 

\section{Top quark decay width}\label{sec:topwidth}
\begin{figure}[htb]
\begin{center}
\includegraphics[width=3in]{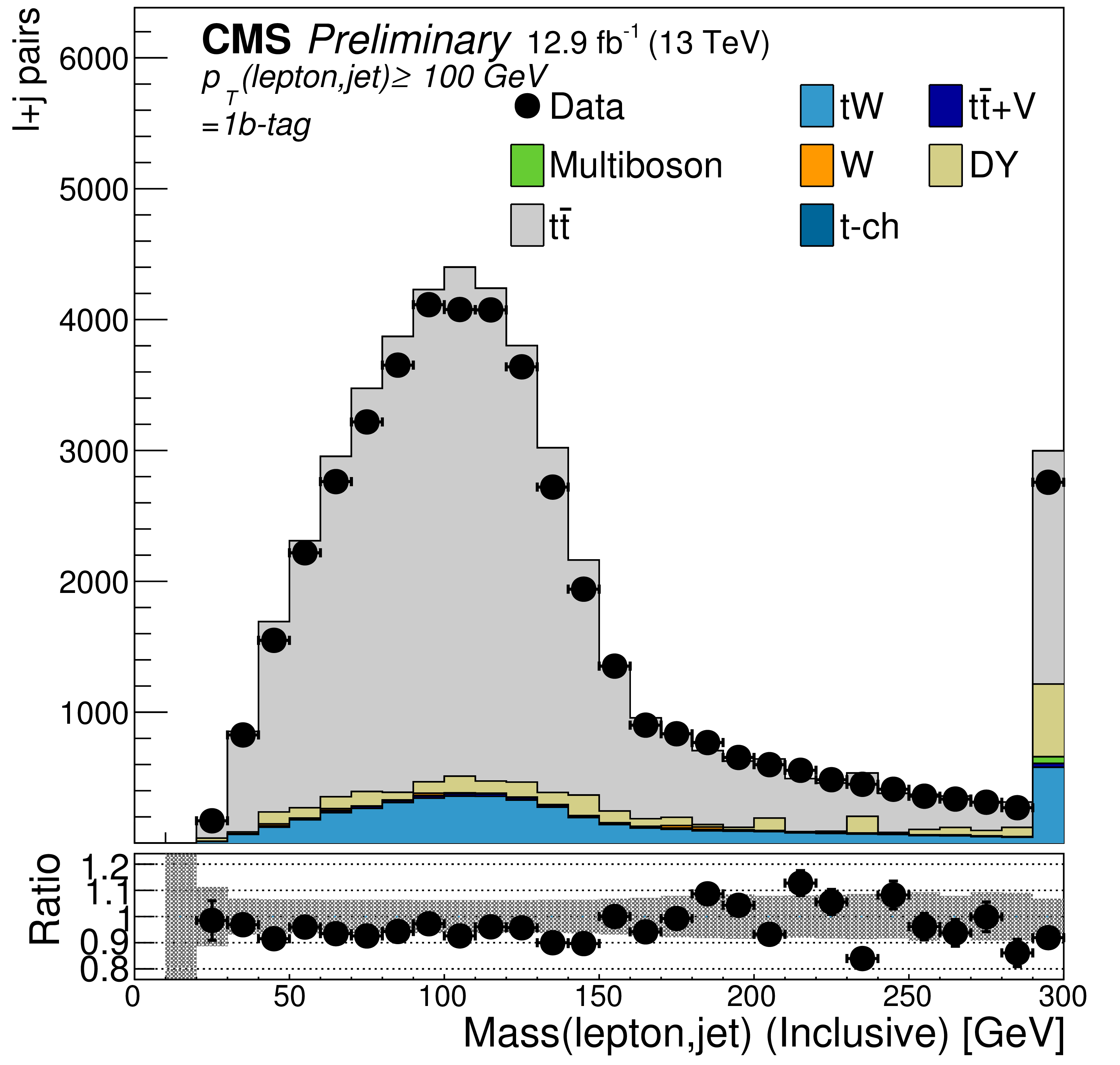}\hfill
\includegraphics[width=3in]{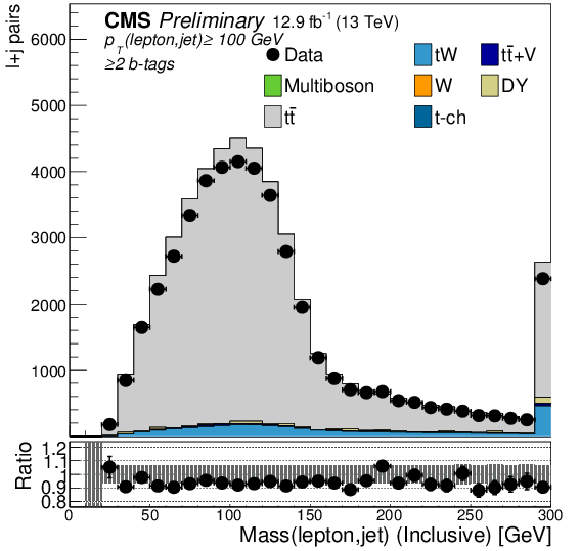}
\end{center}
\caption{Inclusive $\rm M_{lb}$ distributions\cite{ref_decaywidthcms}
in 1 b-tagged (left) and $\rm\geq 2$ b-tagged (right) events where
$\rm p_T^{lb}\geq 100$ GeV.}
\label{fig:cms_topdecaywidth}
\end{figure}

Top quark decay width ($\rm\Gamma_t$) is inversely proportional to its
life-time and any statistically significant deviation from its
SM-predicted value would indicate non-SM decays of top quark.
As presented in Sec.~\ref{sec:topbr}, the CMS measurement for the
${\mathcal R}$ has been translated into an indirect measurement of $\rm\Gamma_t$,
assuming $\rm \sum_q Br(t\rightarrow Wq)$ = 1. The analysis resulted in
$\rm\Gamma_t = 1.36\pm 0.02 (stat)^{+0.14}_{-0.11}$ (syst) GeV, which is in
good agreement with the SM expectations ($\rm\Gamma_{SM}$=1.324 GeV
\cite{ref_SMgamma} for $\rm m_t$=172.5 GeV). Quite recently, CMS has performed a
direct measurement of the same\cite{ref_decaywidthcms} using the
$\rm t\bar{t}\rightarrow$ dilepton events in 13.1 $\rm fb^{-1}$ dataset at
$\rm\sqrt{s}=13$ TeV. The selected events are categorized in terms b-jet multiplicity
(1 b-tagged and $\rm\geq 2$ b-tagged events). The analysis utilizes the
reconstructed invariant mass of the lepton-b-tagged jet systems ($\rm M_{lb}$) as
shown in Fig.~\ref{fig:cms_topdecaywidth}.

In order to extract the limits
on $\rm\Gamma_t$, a two dimensional likelihood fit is performed varying the
signal strength ($\rm\mu=\sigma_{obs}/\sigma_{SM}$) and the sample fraction
for alternative width hypothesis (denoted x). For alternate hypothesis the
the top decay width is considered to have values between $\rm\sigma_{SM}$
and $\rm 4.\sigma_{SM}$, where the signal events are modeled as
\begin{eqnarray*}
\rm N_{signal}=\mu[N_{SM}\cdot (1-x) + N_{alt}\cdot x].
\end{eqnarray*}
Here $\rm N_{signal}$ is the total number of expected signal events
($\rm t\bar{t}$+tW), while $\rm N_{SM}$ ($\rm N_{alt}$) represents
expected numbers of the same in SM (alternate) hypothesis. The observed
and expected limits provide bounds of $\rm 0.6< \Gamma_t< 2.5$ GeV and
$\rm 0.6<\Gamma_t<2.4$ GeV respectively at 95\% confidence level.

\section{Branching Fraction}\label{sec:topbr}
The charge current decays of the top quark i.e., $\rm t\rightarrow Wq$
with q=b, s, or d, are proportional to the corresponding CKM matrix elements
{\it viz.}, $\rm V_{tb},\  V_{ts},\  V_{td}$. However, assuming the unitarity
of CKM matrix and given the measured values for $\rm V_{ub}$ and $\rm V_{cb}$
(or $\rm V_{ts}$ and $\rm V_{td}$), $\rm V_{tb}\approx 1$ and
$\rm |V_{tb}|>> |V_{ts}|,|V_{td}|$. Therefore, a measurement of the ratio of
branching fractions ${\mathcal R}\rm=Br(t\rightarrow Wb)/Br(t\rightarrow Wq)$ with
q=b, s, or d, would be an indirect measure of $\rm |V_{tb}|$. CMS has completed
a measurement of ${\mathcal R}$\cite{ref_topbr} with $\rm t\bar{t}\rightarrow$dilepton+jets
events in 19.7 $\rm fb^{-1}$ of proton-proton collisions dataset at
$\rm\sqrt{s}$= 8 TeV. 

\begin{figure}[htb]
\begin{tabular}{cc}
\includegraphics[width=3.5in]{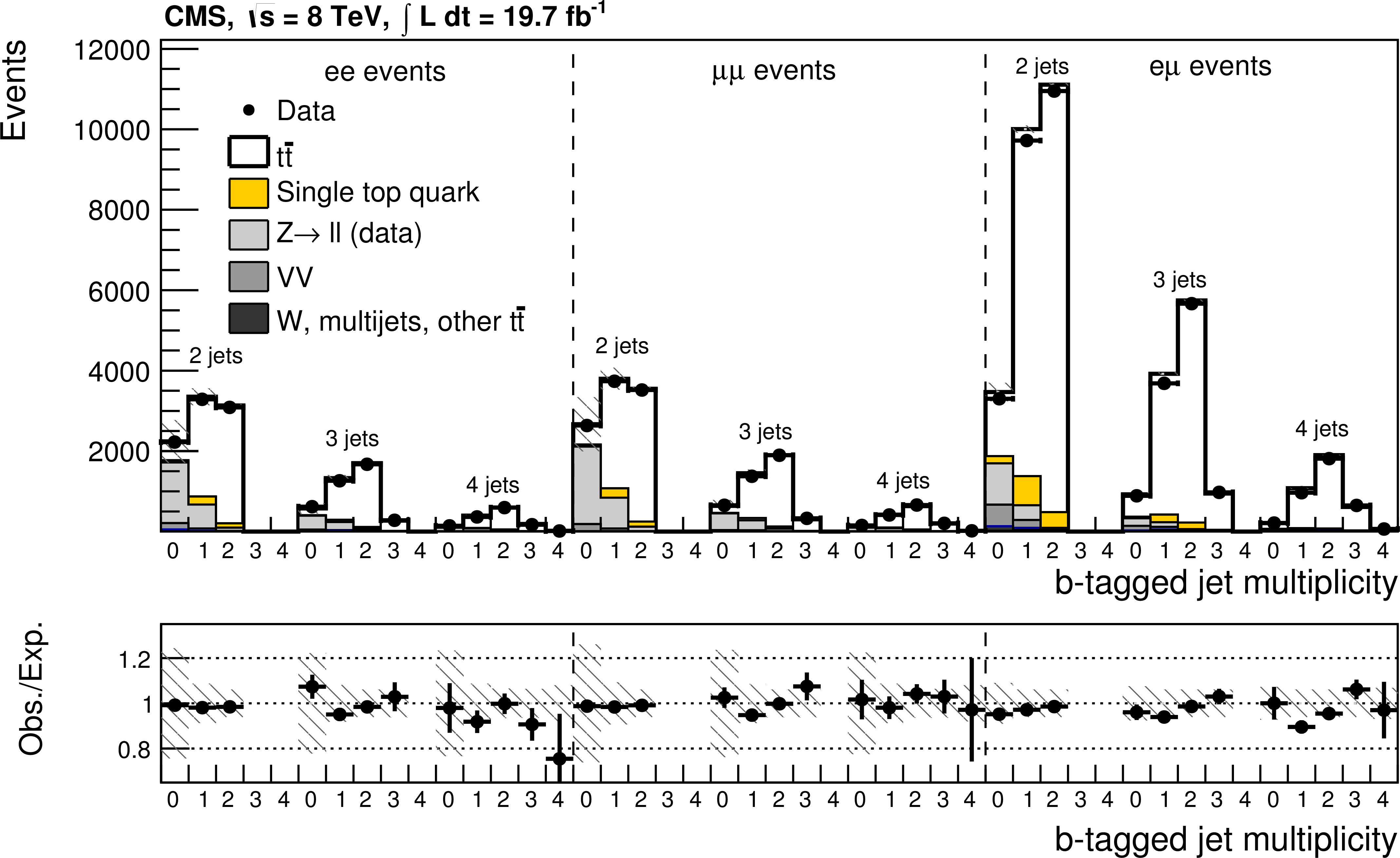}
&\includegraphics[width=2.25in]{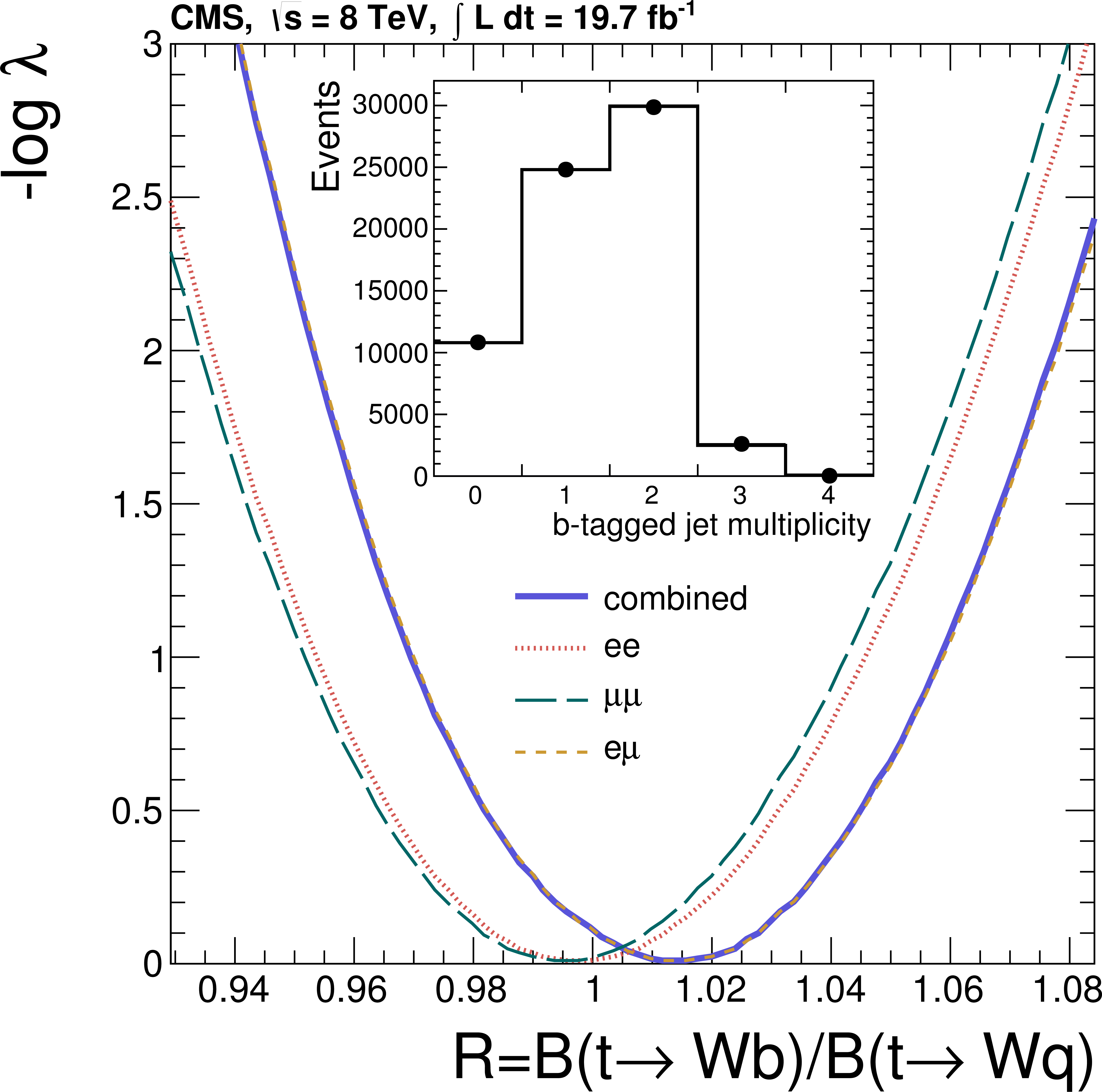}\\
(a)& (b)
\end{tabular}
\caption{(a) Distributions\cite{ref_topbr} of data, SM background and signal events as a function
of b-jet multiplicity in different event category; (b) Profile likelihood
distributions\cite{ref_topbr} as a function of ${\mathcal R}$; the inset shows the inclusive b-tagged jet
multiplicity distribution and the fit distribution.}
\label{fig:cms_topbr}
\end{figure}

In order to have better discrimination against different background processes,
the selection criteria are tuned further based on final state charged lepton
flavors, number of jets and number of b-tagged jets.
Fig~\ref{fig:cms_topbr}(a) shows the agreement between data, SM background and
signal processes ($\rm t\bar{t}$ and tW) in different event categories. The
values of ${\mathcal R}$ are extracted by maximizing the profile likelihood and the results
are displayed in Fig.~\ref{fig:cms_topbr}(b). A combined measurement of ${\mathcal R}
\rm = 1.014\pm 0.003 (stat)\pm 0.032$ (syst) translates to a value of
$\rm |V_{tb}| = 1.007\pm 0.016$ (stat+syst), which is consistent with the SM
expectation. Furthermore, imposing a constraint of ${\mathcal R}\rm \leq 1$ results
in ${\mathcal R}\rm > 0.955$ and $\rm |V_{tb}| < 0.975$ at 95\% confidence level.

\section{Anomalous Wtb coupling}\label{sec:awtb}
\begin{figure}[htb]
\begin{center}
\includegraphics[width=2.1in]{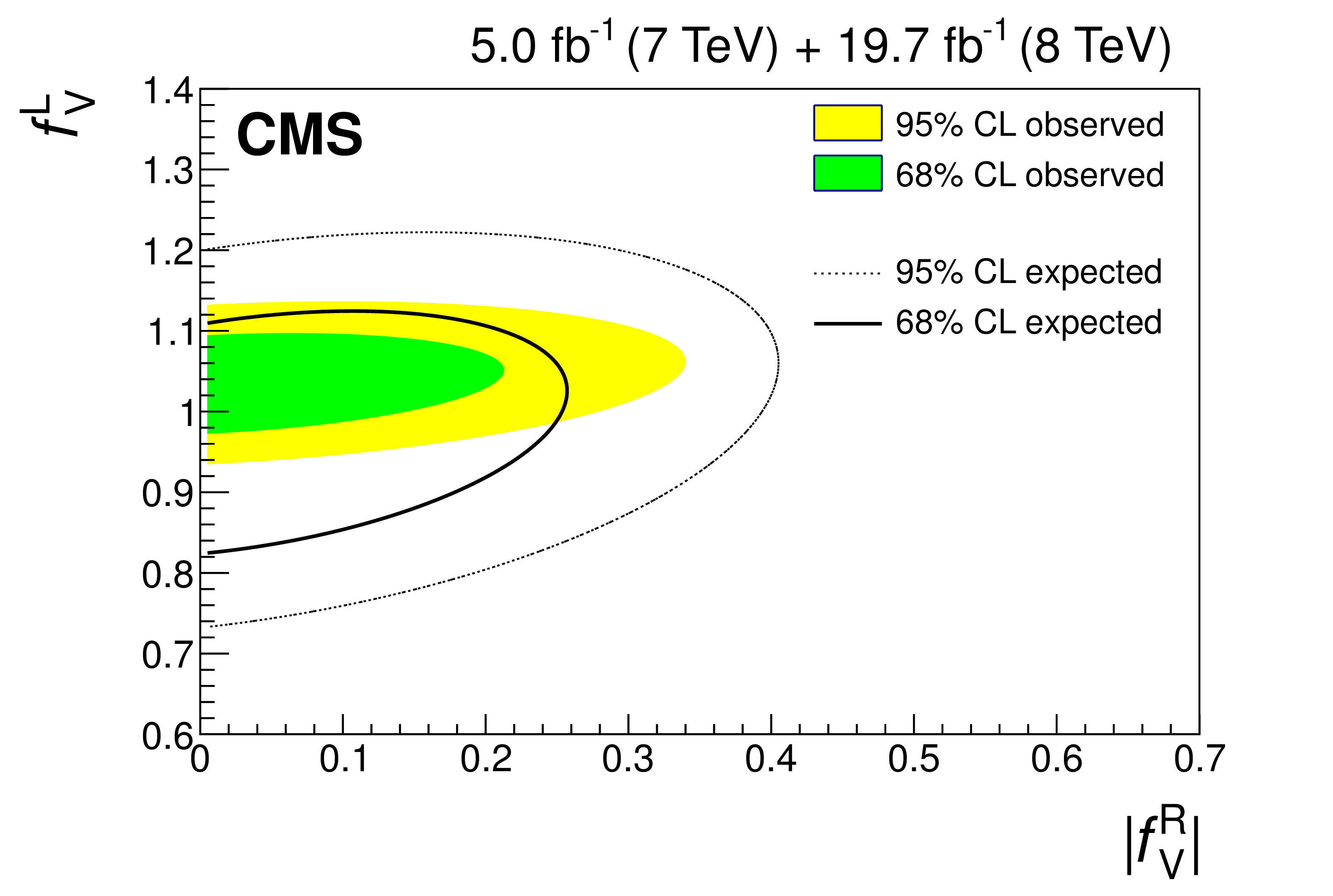}\hfill
\includegraphics[width=2.1in]{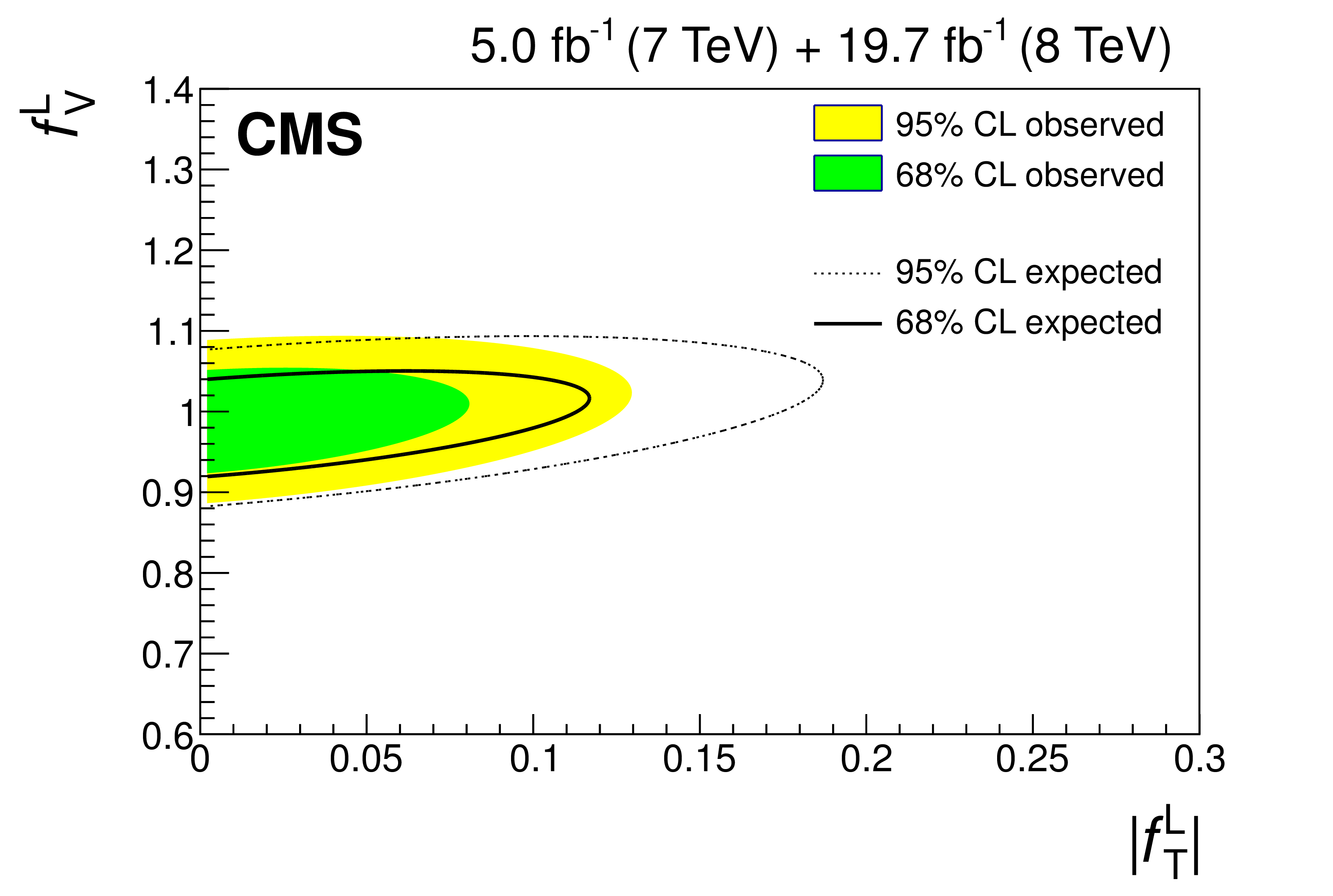}\hfill
\includegraphics[width=2.1in]{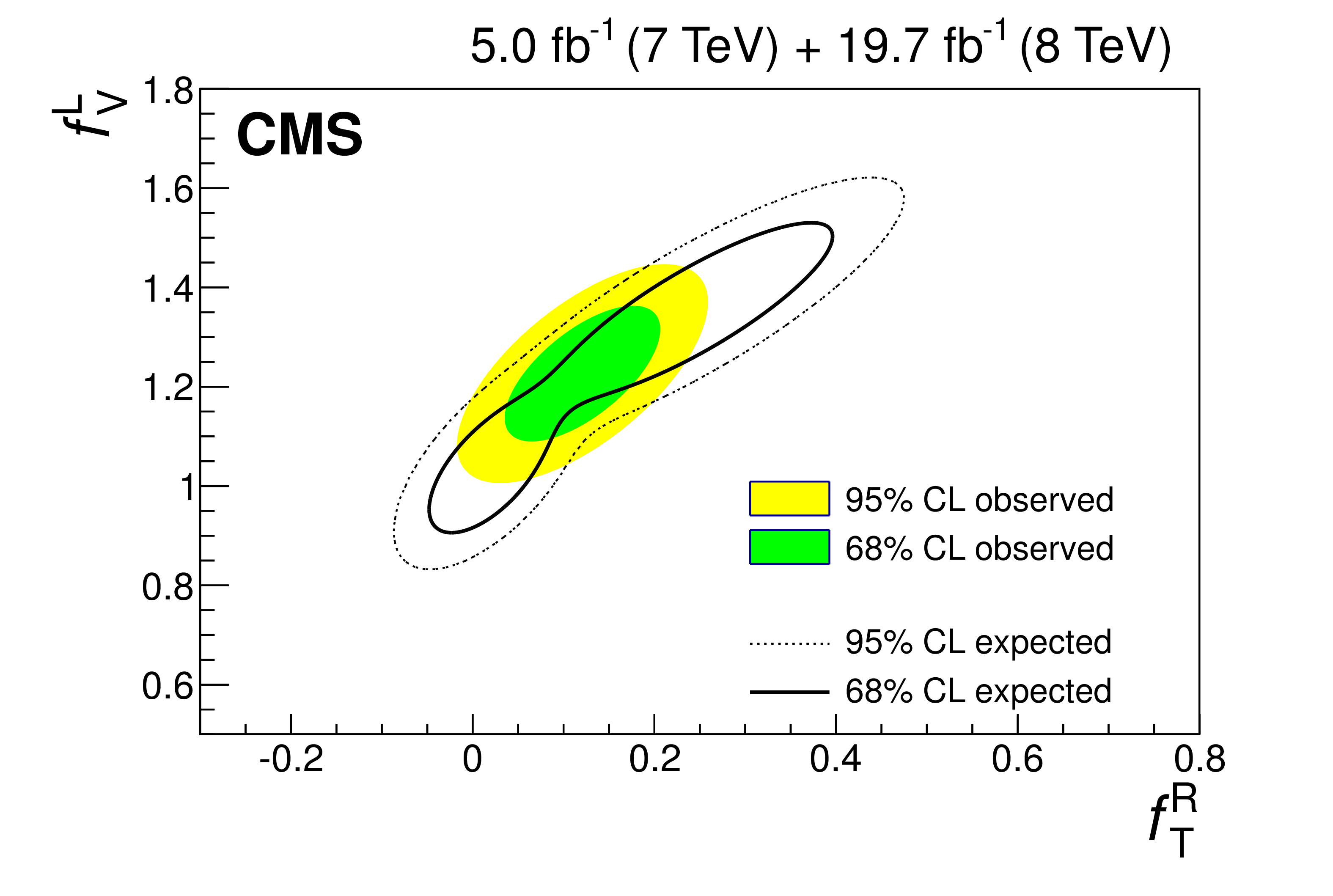}
\end{center}
\caption{ Combined exclusion limits\cite{ref_wtbcms} in the two dimensional planes at
68\% and 95\% confidence levels:
$\rm (f_V^L, |f_V^R|)$ (left), 
$\rm (f_V^L, |f_T^L|)$ (middle), and
$\rm (f_V^L, f_T^R)$ (right).}
\label{fig:cms_wtb}
\end{figure}

The V-A axial-vector structure of the Wtb vertex can be probed directly
through single top production and in particular the t-channel production is quite
sensitive to possible deviations from the SM. The most general CP-conserving
Lagrangian can be written as
\begin{equation}
{\mathcal L} \rm = \frac{g}{\sqrt{2}}\bar{b}\gamma^\mu 
\left( f_V^L P_L + f_V^R P_R \right) tW_\mu^-
 -  \frac{g}{\sqrt{2}}\bar{b}\frac{\sigma^{\mu\nu}\partial_\nu W_\mu^-}{M_W}
\left( f_T^L P_L + f_T^R P_R \right) t +h.c,
\label{eqn:wtb}
\end{equation}
where $\rm f_V^L ( f_V^R)$ and $\rm f_T^L ( f_T^R)$ represent left-handed
(right-handed) vector and tensor couplings respectively; within the SM,
$\rm f_V^L=V_{tb}$ and $\rm f_V^R = f_T^L = f_T^R $ = 0. CMS has completed
a search\cite{ref_wtbcms} in muon + 2 or 3 jets signatures, using the full Run 1 dataset
(5 $\rm fb^{-1}$ at $\rm\sqrt{s}=7$ TeV and 19.7 $\rm fb^{-1}$ at
$\rm\sqrt{s}=8$ TeV). The search utilizes a Bayesian neural network technique
to discriminate between the signal and background processes, which are observed
to be consistent with the SM prediction. The exclusion limits on anomalous
Wtb couplings, {\it i.e.,} $\rm f_V^R$, $\rm f_T^L$ and $\rm f_T^R$ are shown
in Fig.~\ref{fig:cms_wtb} and the observed limits at 95\% confidence level
are set on $\rm |f_V^R| <0.16$,  $\rm |f_T^L|<0.057$ and
$\rm -0.049 <f_T^R <0.048$.

\begin{figure}[htb]
\begin{center}
\includegraphics[width=3in]{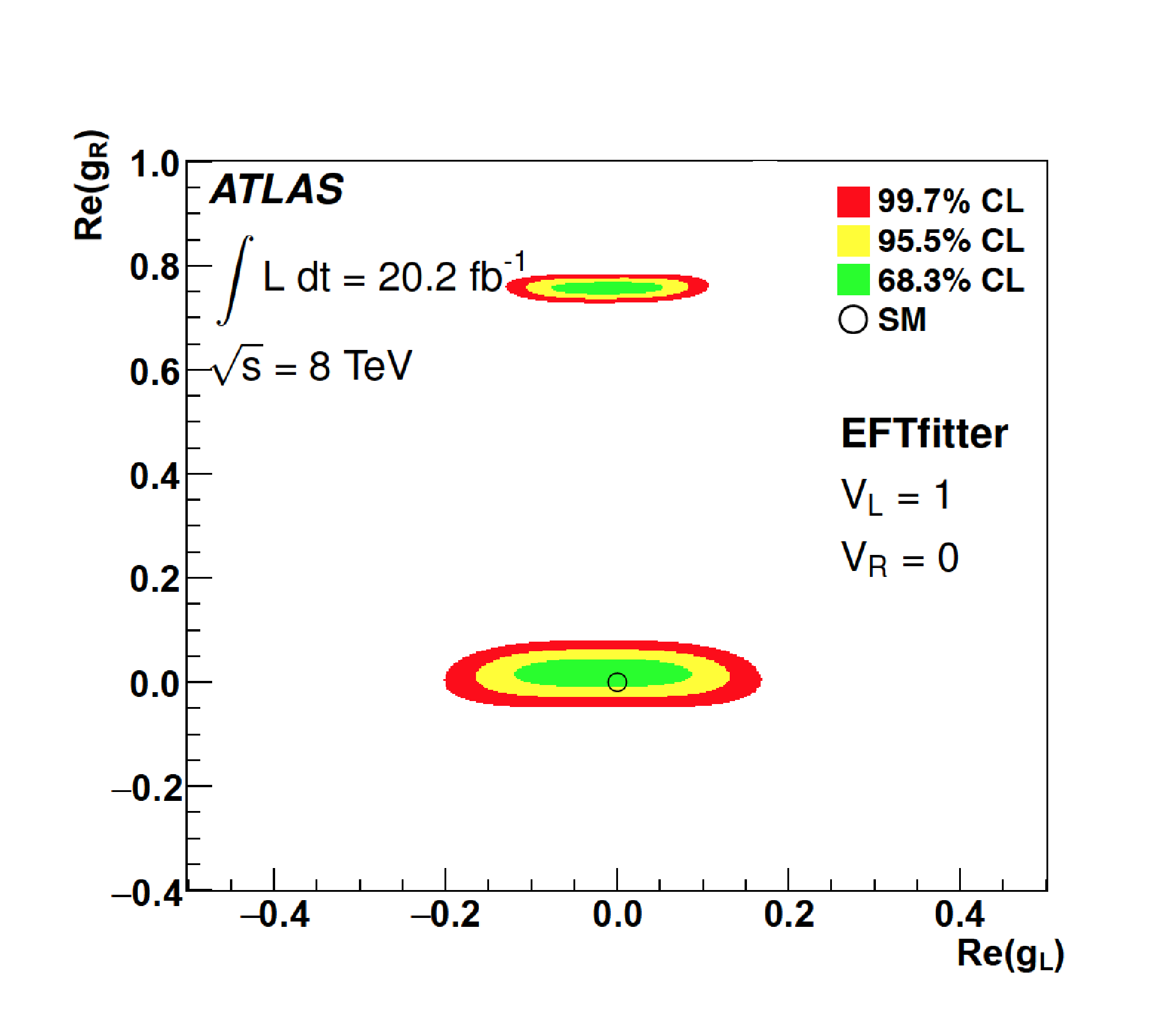}
\includegraphics[width=3in]{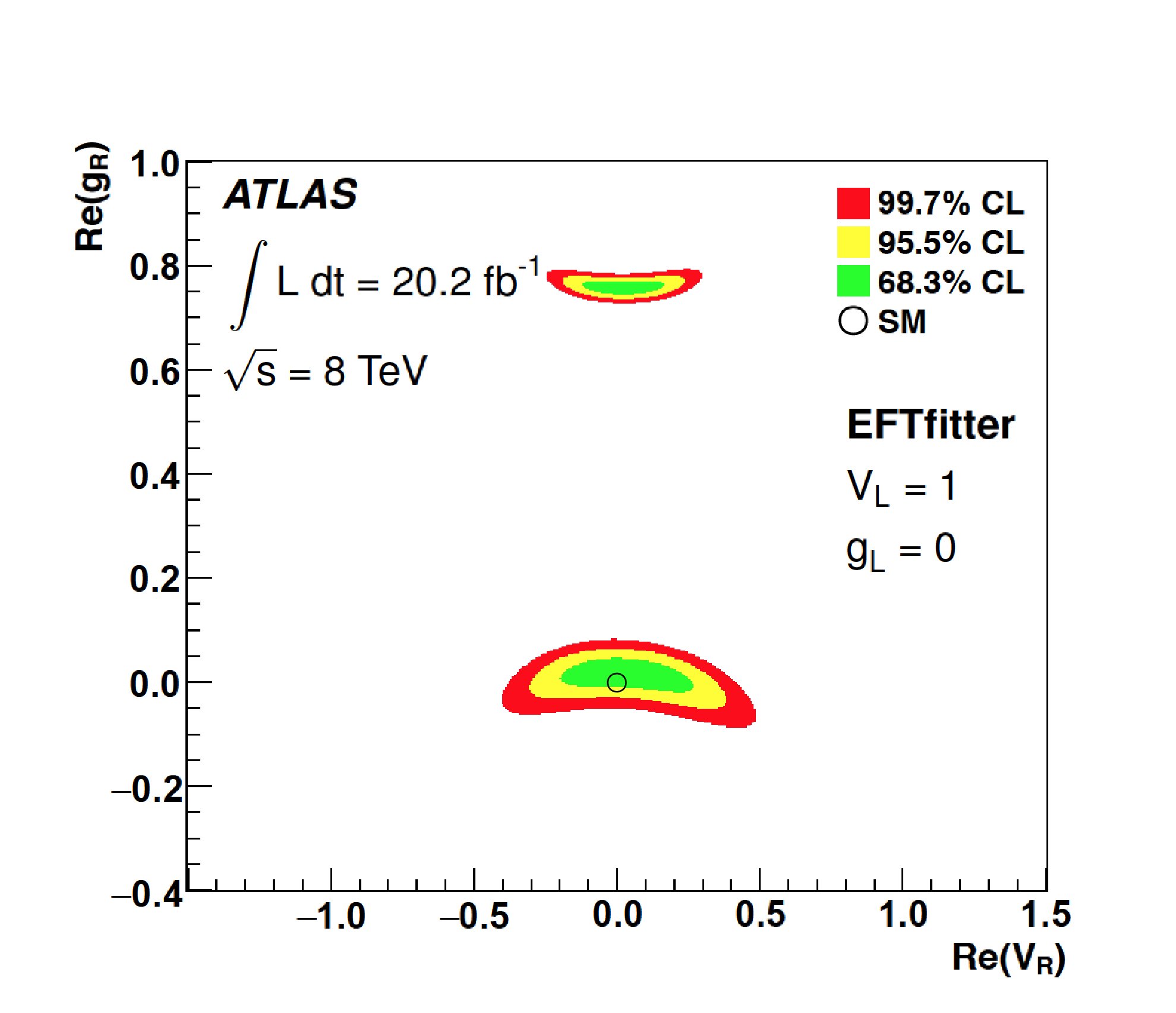}
\end{center}
\caption{ATLAS limits\cite{ref_whelicityatlas} on the anomalous (left) left-handed and right-handed tensor
couplings, and (right) right-handed vector and tensor coupling of the Wtb decay
vertex as obtained from the measured W-boson helicity fractions from the leptonic
analyzer.}
\label{fig:atlas_wtb}
\end{figure}

The ATLAS measurements for the helicity fractions from leptonic analyzer,
(from Sec.~\ref{sec:whelicity}) have also been translated to set limits on
right-handed vector coupling ($\rm V_R$), left-handed tensor coupling ($\rm g_L$)
and right-handed tensor coupling ($\rm g_R$) and the limits are shown in
Fig.~\ref{fig:atlas_wtb}. The observed limits on anomalous $\rm V_R$, $\rm g_L$
and $\rm g_R$ (see Tab.~\ref{tab:atlas_wtb}) are consistent with the SM expectations.

\begin{table}[t]
\begin{center}
\begin{tabular}{c|c}
Couplings &  95\% Confidence Level Interval \\ \hline\hline
$\rm V_R$ & [-0.24,  0.31] \\
$\rm g_L$ & [-0.14, 0.11] \\
$\rm g_R$ & [-0.02; 0:06], [0.74, 0.78]\\ \hline
\end{tabular}
\caption{ATLAS allowed ranges\cite{ref_whelicityatlas} for the anomalous couplings $\rm V_R$, $\rm g_L$ and
$\rm g_R$ at 95\% Confidence Level.}
\label{tab:atlas_wtb}
\end{center}
\end{table}

\section{Conclusions}
Various top quark properties exploring the $\rm t\rightarrow Wb$ decay vertex
have been measured by the ATLAS and CMS experiments using Run 1 and partially
Run 2 datasets. The precision on W-boson helicity fractions in $\rm t\bar{t}$
events is the most accurate till the date and the results are in good agreement
with the SM predictions. CMS direct bounds on top quark decay width are still
limited by statistical uncertainty, while being consistent with other indirect
measurements and with the SM-predicted value of the same. Strong limits on
anomalous Wtb couplings are set excluding any evidence for non-SM decay of
top quark.


\end{document}